\documentclass[12pt]{article}
\usepackage[hmargin=2.5cm,vmargin=3cm]{geometry}
 \usepackage{graphicx}
 \usepackage{amsmath}
 \usepackage{amssymb}
  \usepackage{enumerate}

\begin{document} \title{ Measurements on relativistic quantum fields: II. Detector models  } \author {Charis Anastopoulos\footnote{anastop@physics.upatras.gr} and   Ntina Savvidou\footnote{ksavvidou@upatras.gr}\\
 {\small Department of Physics, University of Patras, 26500 Greece} }

\maketitle

\begin{abstract}
This is the second paper on a new formalism for relativistic quantum measurements.
Here, we construct a fully relativistic model for detectors that takes into account  the detector's  state of motion,  intrinsics dynamics,  initial states and  couplings to the measured field.  The dual classical/quantum description of the detector is implemented by using a master-equation type of approximation for the coarse-grained pointer variables. Then we identify the probabilities that correspond to ideal measurements, i.e., measurements that are largely insensitive to modeling details of the apparatus.  The Unruh-Dewitt and Glauber detectors are recovered at the appropriate limits.
 We employ our results to models of  particle detection, photodetection and relativistic spin measurements, and we derive an ideal distribution for relativistic time-of-arrival probabilities.
\end{abstract}

\section{Introduction}
This is the second in a series of papers on relativistic quantum measurements, presenting a   method   that  applies (i) to any Quantum Field Theory (QFT) and for any field-detector coupling, (ii) for the measurement of any observable, and (iii) for arbitrary size, shape and motion of the detector.

The first paper of the series \cite{AnSav15a} contains the main ideas of the theory including the mathematical set-up and the physical assumptions. Its main result was the derivation of  the probabilities associated to any $n$ distinct measurement events. These probabilities depend on a $2n$-correlation function on the QFT, and also on the physical characteristics of the apparatuses, including their states of motion. All information about each apparatus is contained in a single function, the {\em detector kernel}.

In this paper, we develop a fully relativistic model for  detectors, and we derive a general expression for the detector kernel that is valid in  any set-up. Then, we specialize to three cases of interest: particle detection, photo-detection and relativistic spin measurements. We show that our theory recovers Glauber's \cite{Glauber} and Unruh-Dewitt  detectors \cite{Unruh, Dewitt} at  appropriate limits.

Throughout this work, we employ the word "detector" or "apparatus" to refer to a  detecting element whose records can be correlated with a {\em single microscopic event}. We do not need to assume that the detecting element is elementary in the sense of  \cite{PeTe}, like for instance,
   a single bubble in a bubble chamber or a wire segment in a wire chamber. We assume that the detecting element records only a single event in each run of the particular experiment. In this sense,  an experiment with $n$ distinct  records of observation,  requires $n$    distinct {\em independent} detecting elements, one for each record.

The consideration of {\em temporal coarse-graining}  is essential for the consistent definition of  detection probabilities. In particular, temporal coarse-graining is necessary for a unique determination of the time associated to a measurement event. A record of observation is typically an excitation of finite spatial extension. Different "parts" of the excitation move along different classical trajectories, so they correspond to different proper times. Unless one coarse-grains time at a scale $\sigma$ much larger than the size of the record, the attribution of a time instant to a measurement event is ambiguous. Thus, relativity  sets a lower bound to the scale of temporal coarse-graining characterizing an apparatus. Furthermore,  the coarse-graining time scale also depends on the intrinsic dynamics of the detector and on the effects of the environment.  This dependence is incorporated into a function $\eta$ of the proper time $\tau$, which we call the {\em degradation function}. The degradation function describes the persistence of measurement records in time.

Our modeling of the measurement apparatus leads to  a detection kernel that  is defined through a specification of the following parameters.
\begin{enumerate}[(i)]
 \item an embedding function that describes a detector's world tube in spacetime, i.e., it represents the detector's macroscopic motion;

  \item a set of positive operators describing the measurement of variables other than the spacetime coordinates of the measurement event (e.g., spin, momentum);

   \item the spatial and temporal resolution of the spacetime coordinates; and

   \item the degradation function.
\end{enumerate}

Each choice of the above parameters  corresponds to a physically distinct apparatus. In some regimes, the resulting probability distributions are largely insensitive to fine details in the modeling of the measurement apparatus, These regimes correspond to {\em ideal measurements}. We note that, with this definition,  ideal measurements are not, in general, projective measurements.

After deriving a general expression for the detection kernel, we apply  the formalism to specific types of measurement. First, we derive an ideal probability distribution for the time-of-arrival of relativistic particles. This probability distribution generalizes the relativistic POVM of Ref. \cite{AnSav12} and is compatible with different couplings of the field to the measurement apparatus. We also consider the case of photodetection, where we show that Glauber's theory is obtained at the limit that $\sigma$ is much smaller than any time-scale characterizing the state of the electromagnetic (EM) field. Finally, we derive  a general expression the time-integrated probability density associated to relativistic spin measurements.
\medskip

This paper is structured as follows. In Sec. 2, we briefly summarize the results of Ref. \cite{AnSav15a}, setting the background for the present work. In Sec. 3, we present an important approximation for the self-dynamics of the internal degrees of freedom of the apparatus. In Sec. 4. we present our  model for the measurement apparatus, and derive a workable expression for the detector kernel. This is employed in Sec. 5, in order to derive a general characterization of relativistic detectors. In the following three sections, we present applications of the formalism: particle detection (Sec. 6), photo-detection (Sec. 7) and spin measurements (Sec. 8). In Sec. 9, we summarize and discuss our results.

 \section{Summary of past work}

The main result of Ref. \cite{AnSav15a} is the construction of a probability density associated to $n$ measurement events. The physical system under consideration is described by a QFT on Minkowski spacetime. The quantum  fields $\hat{\phi}_r(X)$ are defined on a Hilbert space ${\cal F}$ and they transform covariantly under a representation of the Poincar\'e group on ${\cal F}$. The fields interact with $n$ independent detectors, resulting to $n$ independent measurement records.

\medskip

{\em Spacetime description of detectors.} Each detector admits a dual-quantum classical description.
The spatial extension of the detector at rest corresponds to a subset $S$ of Euclidean space ${\pmb R}^3$. We represent the points of $S$ by three-vectors ${\pmb q}$.
A moving detector corresponds to a world-tube
 $S \times {\pmb R}$ in Minkowski spacetime,  described by a spacelike embedding function

\begin{eqnarray}
{\cal E}: S \times {\pmb R} \rightarrow M. \label{embed}
\end{eqnarray}

 The embedding is expressed in terms of coordinate functions ${\cal E}^{\mu}(\tau, {\pmb q})$, for $\tau \in {\pmb R}$ and ${\pmb q} \in S$. Given the embedding function, we define the frame vector fields on the world-tube
 \begin{eqnarray}
 \dot{\cal E}^{\mu} = \frac{\partial {\cal E}^{\nu}}{\partial \tau}  \hspace{1cm} {\cal E}^{\mu}_i = \frac{\partial {\cal E}^{\nu}}{\partial q^i}.
 \end{eqnarray}

 The time-variable $\tau$ is  the proper time
  of the path ${\cal E}^{\mu}(\tau, q)$ for fixed ${\pmb q}$, i.e., so that $\eta_{\mu \nu}  \dot{\cal E}^{\mu}  \dot{\cal E}^{\nu} = -1$. We assume that each measurement event is specified by an accuracy of order $\sigma$ in the proper time; $\sigma$ is the temporal coarse-graining scale for the apparatus. The, the detector's size  must be much smaller than $\sigma$, so that a {\em single proper time parameter} suffices for the temporal ordering of recorded events.

\medskip

{\em Quantum description of detectors.}  Each detector is associated to a Hilbert space ${\cal K}$ that splits into two complementary subspaces: ${\cal K} = {\cal K}^+ \oplus {\cal K}^-$.
 The subspace  ${\cal K}^-$   corresponds to the absence and  the subspace ${\cal K}^+$  to the presence of a macroscopic measurement record.  We denote the projection operator associated to ${\cal K}_+$ by $\hat{E}$, and assume an initial state $|\omega \rangle \in {\cal K}_-$.

Consider, for example,  an avalanche diode.  ${\cal K}$  represents the Hilbert space of holes and electrons in a junction, subject to suitable boundary conditions (external voltage, temperature and so on). Let $\hat{I}$ be an operator that corresponds to the current through the diode. Then, ${\cal K}_-$ is the subspace corresponding to all eigenvalues of $\hat{I}$ smaller than some critical value $I_0$ (significantly larger than the current due to thermal fluctuations) and ${\cal K}_+$  corresponds to all eigenvalues of $\hat{I}$ larger than $I_0$.

 The self-dynamics of a detector is described  by a Hamiltonian operator $\hat{h}$ on ${\cal K}$. The Hamiltonian $\hat{h}$ generates translations with respect to the proper time $\tau$. The self-dynamics of the evolution operator ought to leave the subspaces ${\cal K}_{\pm}$ invariant, otherwise it would continuously generate measurement records, even in absence of a measured system. Hence, $[\hat{E}, \hat{h}] = 0$.

The  interaction Hamiltonian describing the coupling between field and detector is

\begin{eqnarray}
 \hat{V}= \int d^3 x \hat{Y}_A({\pmb x})   \hat{J}^A({\pmb x})
  \label{vterm}
\end{eqnarray}
 where $\hat{Y}_A({\pmb  x})$ is a composite operator on ${\cal F}$ that is a local functional of the fields $\hat{\phi}_r$, and $\hat{J}^A({\pmb x})$ are current operators  defined on the Hilbert space ${\cal K}$ of the   detector. $A$ is a collective index for the composite operators.

 \medskip

{\em Observables.}
Let $\lambda$ stand for all observables recorded by  the  detector. We denote the associated  positive operators on  ${\cal K}$ by $\hat{F}(\lambda)$ . In absence of a detection event, there are no observables so the operators $\hat{F}(\lambda)$ are exhaustive on the subspace ${\cal K}^+$ associated to detection,
  $\sum_{\lambda} \hat{F}(\lambda) = \hat{E}$.

 The set of measurement outcomes always includes the location of the detection event. Hence, the parameter $\lambda$ in the positive operators $\hat{F}(\lambda)$ is a shorthand for the pair $({\pmb Q}, \mu)$ where ${\pmb Q}$ is a pointer variable that correlated to the coordinate ${\pmb q}$ of the detector, and $\mu$ refers to pointer variables for magnitudes other than position (e.g., momentum and spin).

Both  $\mu$ and ${\pmb Q}$ are highly coarse-grained variables, since they corresponds to macroscopic records. Hence, $\hat{F}({\pmb Q}, \mu)$ can be expressed as a product
 \begin{eqnarray}
\hat{F}({\pmb Q}, \mu) =  \hat{F}_1({\pmb Q}) \hat{F}_2(\mu), \label{ff1f2}
 \end{eqnarray}
 where $\hat{F}_1({\pmb Q})$ and $\hat{F}_2(\mu)$ are POVMs for the pointer variables ${\pmb Q}$ and $\mu$ \cite{Omn88}.   This implies that
$\int d^3Q \hat{F}_1({\pmb Q}) = \hat{E}$ and $\sum_{\mu} \hat{F}_2(\mu) = \hat{E}$.

\medskip

{\em Stationarity condition.} We assume that the initial state $|\omega \rangle$ of the detector satisfies the {\em stationarity condition}
\begin{eqnarray}
\hat{J}^A({\pmb x}) e^{-i \hat{h} \tau} |\omega\rangle = \int d^3q \hat{J}^A({\pmb q}) \delta^3(x^i - {\cal E}^i ({\pmb q}, \tau))|\omega'\rangle, \label{jcu}
\end{eqnarray}
for some vector $|\omega'\rangle$ and current operators $\hat{J}^A({\pmb q}) $ defined on $S$.
The stationarity condition (\ref{jcu}) implies that
 the current operators and the initial state are combined in such a way such that the only time dependence  of $\hat{J}^A({\pmb x}, \tau) e^{-i \hat{h} \tau}|\omega\rangle $ is due to the motion of the apparatus.

 For a static detector, Eq. (\ref{jcu})   means that $|\omega \rangle$ is an eigenstate of $\hat{h}$, and the corresponding energy eigenvalue is  chosen to be zero.
  For a moving detector, Eq. (\ref{jcu}) means that the Hamiltonian  $\hat{h}$ affects only the part of the quantum state that corresponds to the apparatus'    macroscopic motion.

{\em Probabilities.} Given the assumptions above, we compute the probability density for a sequence of $n$ measurement events by $n$ detectors. Each event is characterized by a value of the proper time variable $\tau$ of the associated detector,
one value for the observable ${\pmb Q}$ that is correlated to the position, and other observables $\mu$. It is important to emphasize that both the time and space coordinates are treated as random variables, and the probabilities define densities with respect to time.
This is achieved using  the Quantum Temporal Probabilities (QTP) method  \cite{AnSav12} that has been developed for addressing   the time-of-arrival problem in quantum mechanics \cite{AnSav12, AnSav06}.

In the QTP method, the time of an event is treated as a quasiclassical macroscopic variable. For such variables,  probabilities can be rigorously defined. This implies that we must introduce a coarse-graining  time scale $\sigma$ that corresponds to the minimal temporal resolution of any event recorded by the detector. In the present context, the scale $\sigma$ refers to proper time rather than coordinate time. We shall explain the origins and physical interpretation of $\sigma$ in the next section.

In Ref. \cite{AnSav15a}, we computed the probability density  associated to $n$ detection events, each corresponding to a specific value of the proper time $\tau_i$, of the coordinate ${\pmb Q}_i$ and the observable $\mu_i$ of the $i$-th detector,
\begin{eqnarray}
P(\tau_1, {\pmb Q}_1, \mu_1;\tau_2, {\pmb Q}_2, \mu_2;\ldots; \tau_n, {\pmb Q}_n, \mu_n ) =  \nonumber \\  \sum_{A_1, \ldots, A_n} \sum_{B_1, \ldots, B_n} \int d^4X_1 \ldots dX_n \int d^4X'_1 \ldots dX'_n
G(X_1, A_1; \ldots, X_n, A_n| X_1', B_1; \ldots; X_n', B_n)\nonumber \\
\times R_{(1)}(X_1,A_1; X_1', B_1| \tau_1, {\pmb Q}_1, \mu_1) \ldots R_{(n)}(X_n,A_1; X_n', B_1| \tau_n, {\pmb Q}_n, \mu_n),  \label{longh}
 \end{eqnarray}
where $i = 1, 2, \ldots, n$ labels the apparatuses. The correlation function  $G$ in Eq. (\ref{longh}) has $2n$ arguments and is defined as
\begin{eqnarray}
 G(X_1, A_1; \ldots, X_n, A_n| X_1', B_1; \ldots; X_n', B_n) = \nonumber \\
 Tr
\left[T[\hat{Y}_{A_n}(X_n) \ldots \hat{Y}_{A_1}(X_1)] \hat{\rho}_0  \bar{T}[\hat{Y}_{B_1}(X') \ldots \hat{Y}_{B_n}(X'_n)] \right], \label{2np}
\end{eqnarray}
 where   $\hat{\rho}_0$ is the initial state of the quantum field,    $X = ({\pmb x}, t)$ and $\hat{Y}_A(X) = \hat{Y}_A({\pmb x}, t) = e^{i\hat{H}_{\Phi}t} \hat{Y}_A({\pmb x}) e^{-i\hat{H}_{\Phi}t}$ is the Heisenberg picture  composite operator of the interaction Hamiltonian.  We note that the correlation function $G$ has $n$ time-ordered arguments ($T$) and $n$ arguments in reversed time order ($\bar{T}$).

All information about the $i$-the detector is contained in the function $R_{(i)}$, the {\em detector kernel}, which is defined as follows.

\begin{eqnarray}
R(X, A; X', B| \tau, {\pmb Q}, \mu) &=& \int ds g_{\sigma}(s) \int_S d^3q \int_S d^3q'  \delta^4[ X - {\cal E}(\tau +\frac{s}{2}, {\pmb q})]  \delta^4[ X' - {\cal E}(\tau -\frac{s}{2}, {\pmb q}')] \nonumber
\\
&\times& \langle \omega'| \hat{J}^{B}({\pmb q}') \sqrt{\hat{F}}({\pmb Q}, \mu)
e^{i\hat{h}s} \sqrt{\hat{F}}({\pmb Q}, \mu)  \hat{J}^{A}({\pmb q})|\omega'\rangle. \label{kernel2}
\end{eqnarray}
In Eq. (\ref{kernel2}), $g_{\sigma}$ is a function of proper time which can be conveniently chosen as a Gaussian $g_{\sigma}(s) = \exp[-s^2/(8\sigma^2)]$.

In the following two sections, we will elaborate on Eq. (\ref{kernel2}) and derive a simple equation for the detector kernel.

\section{The microscopic origins of the coarse-graining scale}

\subsection{The master-equation approximation}
The probability density (\ref{longh}) is well defined provided that the temporal variables are averaged at a time-scale $\sigma$. The  scale $\sigma$ corresponds to the {\em irreducible} temporal resolution of the apparatus. By "irreducible", we mean that it arises out of the fundamental quantum description of the apparatus. The actual resolution of an apparatus also includes contributions from classical uncertainties, for example, uncertainties in the propagation of the signal and uncertainties of the classical clock that determines the timing of an event. In what follows, we explain how $\sigma$ is inferred from the intrinsic dynamics of the apparatus.

First, we recall von Neumann's model of quantum measurements \cite{vNeu}.  In this model, the self-Hamiltonian of the apparatus is negligible when compared to the  interaction between the apparatus and the microscopic system. The interaction term causes the correlation between the microscopic observable and the pointer variable of the apparatus. One expects that a small but non-zero self-Hamiltonian induces fluctuations of the pointer variable around the recorded value, i.e., a degradation of the measurement signal. In our model, we provide a characterization of this degradation.

Let us use the  symbol $\lambda$ to denote all measurement records ${\pmb Q}$ and $\mu$ in Eq.  (\ref{kernel2}). The records are represented by the positive operators $\hat{\Pi}(\lambda)$. They correspond to  {\em macroscopic} pointer variables that are defined in terms of collective  degrees of freedom in the apparatus.  This means that the operators $\hat{F}(\lambda)$ describe highly coarse-grained observables.

There are several different ways for quantifying the  coarse-graining of a positive operator $\hat{F}(\lambda)$. For example, we can define   the associated density matrix
 \begin{eqnarray}
\hat{\rho}_{\lambda} = \frac{\hat{F}(\lambda)}{Tr\hat{F}(\lambda)}, \label{rhola}
\end{eqnarray}
  and use the   von Neumann entropy $S_{\lambda}  = -Tr (\hat{\rho}_{\lambda} \log \hat{\rho}_{\lambda})$ as a measure of coarse-graining.

 In Eq. (\ref{kernel2}), the  detector's self-Hamiltonian $\hat{h}$ appears in a term
$ \sqrt{F}(\lambda)e^{-i \hat{h}s} \sqrt{F}(\lambda)$.  We evaluate this product using   the fact that $\hat{F}(\lambda)$ is a highly coarse-grained operator. To this end, we adapt an approximation that is commonly employed
   in the derivation of master equations for macroscopic variables \cite{OPenr}.   We write
 \begin{eqnarray}
 \sqrt{F}(\lambda)e^{-i \hat{h}s} \sqrt{F}(\lambda)  = \hat{F}(\lambda) \eta(\lambda, s) \label{mastereq},
 \end{eqnarray}
 where   the function
 \begin{eqnarray}
 \eta(\lambda, s) = \frac{Tr \left[ \hat{F}(\lambda)e^{-i\hat{h}s}\right]}{Tr\hat{F}(\lambda)} \label{etadef}
 \end{eqnarray}
  will be referred to as the {\em degradation function} of the detector\footnote{If the operators $\hat{F}(\lambda)$ are not trace class, a suitable regularization is necessary before taking the quotient in Eq. (\ref{etadef}).}.     Eq. (\ref{mastereq}) is the {\em master-equation approximation} for the pointer variables.
 For analogous conditions of classicality for coarse-grained quantum variables, see Refs. \cite{Omn88, Omn95, Ana95}.

For a justification of Eq. (\ref{mastereq}), let us first consider the case of projective
  $\hat{F}(\lambda)$, i.e.,   assume that $\sqrt{\hat{F}(\lambda)}  =  \hat{F}(\lambda)$ and $\hat{F}(\lambda) \hat{F}(\lambda') = 0$, for $\lambda \neq \lambda'$. For each $\lambda$,  $\hat{F}(\lambda)$ projects into  a subspace ${\cal K}_{\lambda}$  of the Hilbert space ${\cal K}_+$.

  Since $\lambda$ corresponds to macroscopically distinguishable observables, there is a conceptual analogy between the subspaces ${\cal K}_{\lambda}$ and the  macrostates of non-equilibrium statistical mechanics. In systems with a large number of degrees of freedom, the microscopic quantum dynamics are expected to generate an effective stochastic dynamics at the level of the macrostates.  That is, we expect that the Hamiltonian evolution $e^{-i\hat{h}s}$ induces a probabilistic evolution between the subspaces ${\cal K}_{\lambda}$ characterized by different values of $\lambda$. The latter is expressed in terms of a transition probability function $G(\lambda, 0; \lambda', s)$ expressing the conditional probability that the recorded variable will take value $\lambda'$ at time $s$, provided it took value $\lambda$ initially.

In non-equilibrium statistical mechanics, the transition probability $G(\lambda, 0; \lambda', s)$ is commonly estimated via a {\em rerandomization} condition, like Boltzmann's {\em Stosszahlansatz}, or the transfer matrix approximation \cite{sklar}. Here, we assume  that for sufficiently large times $s$, the density matrix of the system is well approximated by the density matrix (\ref{rhola}) of maximum ignorance on ${\cal K}_{\lambda}$.
 Then,
\begin{eqnarray}
G(\lambda, 0; \lambda', s) = Tr (\hat{F}(\lambda') e^{-i\hat{h}s}\hat{\rho}_{\lambda} e^{i\hat{h}s}) = \frac{Tr \left[    \hat{F}(\lambda') e^{-i\hat{h}s} \hat{F}(\lambda) e^{i \hat{h}s}  \right]}{Tr \hat{F}(\lambda)} \label{mastereq2}
\end{eqnarray}
 Eq. (\ref{mastereq2}) is an analogue of Pauli's master equation, in the sense that it monitors only the probabilities with respect to the different subspaces ${\cal K}_{\lambda}$ and ignores cross-terms in the density matrix.

The rerandomization assumption implies that the evolution operators   are assumed to act only on states of the form $\hat{\rho}_{\lambda}$ and produce states  of the form $\hat{\rho}_{\lambda'}$ with some multiplicative coefficient, i.e., it implies that
\begin{eqnarray}
\hat{F}(\lambda)e^{-i \hat{h}s}\hat{F}(\lambda) = \eta(\lambda, s) \hat{F}(\lambda),  \label{mastereq0}
\end{eqnarray}
 with $\eta(\lambda, s)$ given by Eq. (\ref{etadef}).
Substituting into Eq. (\ref{mastereq2}), we find
\begin{eqnarray}
G(\lambda, 0; \lambda, s) = |\eta(\lambda, s)|^2.
\end{eqnarray}

Eq. (\ref{mastereq0}) was justified for projective $\hat{F}(\lambda)$. Continuity implies that its analogue also holds  for approximate projectors, i.e., for $\hat{F}(\lambda)$ that satisfy the conditions
\begin{eqnarray}
\frac{Tr| \hat{F}(\lambda) - \sqrt{\hat{F}(\lambda)}|}{ Tr \hat{F}(\lambda)} << 1, \hspace{1cm} \frac{Tr| \hat{F}(\lambda) \hat{F}(\lambda')|}{ Tr \hat{F}(\lambda)} << 1 \hspace{0.3cm} \mbox{if} \hspace{0.2cm} \lambda \neq \lambda'.
\end{eqnarray}
Then Eq. (\ref{mastereq}) is obtained.

\subsection{Meaning  of the degradation function}

It is easy to show that the degradation function  satisfies
\begin{eqnarray}
|\eta(\lambda, s)| \leq 1 \\
\eta(\lambda, 0) = 1.
\end{eqnarray}

The degradation function  $\eta(\lambda, s)$ coincides with  the {\em persistence amplitude} \cite{FGR78, CSM77, Per80} of the  density matrix $\hat{\rho}_{\lambda}$, Eq. (\ref{rhola}).  The square norm $|\eta(\lambda, s)|^2$ is the associated persistence probability.
  We chose the characterization "degradation" rather than "persistence" for    $\eta(\lambda, s)$ because we want to emphasize the degradation of the record due to the intrinsic dynamics of the apparatus, rather than the persistence of some initial state.

 The squared norm of the degradation function  is expected to be a decreasing function of $s$.  This is a typical behavior  for persistence probabilities \cite{Per80}; in fact, the decay is often exponential. This is consistent with the fact that  the evolution generated by Eq. (\ref{mastereq2}) is typically diffusive, and `dilutes'  any initial concentration of the probability distribution at a specific value of $\lambda$.

 Let us denote by $\tau_d$ the time scale characterizing the decay of $\eta(\lambda, s)$. Eq. (\ref{kernel2}) implies that if $\tau_d$ is much larger than the temporal coarse-graining time-scale $\sigma$, the degradation function does not influence the probabilities:
  for $s < \sigma$, $\eta(\lambda, s) \simeq \eta(\lambda, 0) = 1$, while all contributions to the detector kernel  from $s >> \sigma$ vanish due to the function $g_{\sigma}$. In contrast, for $\sigma >> \tau_d$, $g_{\sigma}$ is effectively unity.
  The key point is that the degradation function suppresses all interferences from amplitudes with  detection times  differing by $\delta t >> \tau_d$. Thus, choosing $\sigma$ of $\tau_d$ or larger guarantees the consistent definition of probabilities---see, Sec. 2.2 in Ref. \cite{AnSav15a}.

   We conclude that the optimum value for $\sigma$ that is compatible with the definition of probabilities and leads to maximal resolution in the time measurements is of the order of $\tau_d$. Thus, {\em the degradation function of the detector naturally determines the scale of temporal coarse-graining}.

For many observables, different values of the pointer variable do not correspond to different energies. For example, in a homogenous detector, there is no energy difference between detection records at different points. In such cases, the degradation function  is insensitive to the specific value $\lambda$, and we denote it as $\eta(s)$.

 In what follows, we estimate the degradation function and the associated timescale $\tau_d$ for  two classes of measurements.

 \subsection{Energy measurements.} We consider an apparatus in which the recorded variable is the deposited energy, i.e.,  a calorimeter\footnote{A macroscopic record of energy absorption typically appears   within an elementary detecting element, i.e., it corresponds to a transition in the energy levels of a single detecting element. Whether one focuses on the energy of a single element or of the whole detector is not an issue, because energy is an extensive variable. However, when trying to ascertain how the number $N$ of particles in the detector is related to the decay scale $\tau_d$, it is important to keep in mind that $N$ refers to the number of particles of an elementary detector.}. The positive operators associated to energy measurements are

\begin{eqnarray}
\hat{F}(E) = \int dE' f_{\Delta}(E - E') \hat{P}_{E'},
\end{eqnarray}
where $\hat{P}_E$ are the spectral projectors of the Hamiltonian $\hat{h}$. The sampling function  $f_{\Delta}(E)$ is localized around $E = 0$ with width $\Delta$ and satisfies $\int dE f_{\Delta}(E) = 1$.

 Then, Eq. (\ref{etadef}) gives
 \begin{eqnarray}
 \eta(E,s) = \frac{ \int dE f_{\Delta}(E - E') e^{-iE's} g(E')}{\int dE f_{\Delta}(E-E') g(E')},
 \end{eqnarray}
where $g(E)$ is the density of states of the apparatus.

We write $g(E) = e^{S(E)}$, and we expand
\begin{eqnarray}S(E') = S(E) + \frac{1}{T(E)} (E'-E) - \frac{1}{2C(E) T(E)^2}  (E'-E)^2.
 \end{eqnarray}
The functions  $S(E)$,  $T^{-1}(E) = \partial S/\partial E$ and $C^{-1}(E)  = \frac{\partial T}{\partial E}$ can be identified with the entropy,  the temperature and the heat capacity of the microcanonical distribution associated to $\hat{h}$, respectively.

For energies $E >> \Delta$, we can use  a Gaussian sampling function,
$f_{\Delta}(E) = \frac{1}{\sqrt{2\pi \Delta^2}} e^{- \frac{E^2}{2\Delta^2}}$, to obtain

\begin{eqnarray}
\eta(E,s) = \exp \left[ -\frac{s^2}{2[\Delta^{-2} + (CT^2)^{-2}]}- is \left(E + (\frac{T}{\Delta^2} + \frac{1}{CT}\right)^{-1}     \right].
\end{eqnarray}
In the regime where the sampling accuracy is much larger than the thermal fluctuations of the energy, $\Delta >> \sqrt{C} T$, the degradation function simplifies
\begin{eqnarray}
\eta(E,s) = e^{-iEs} e^{- \frac{CT^2}{2}s^2}.
\end{eqnarray}
The decay time-scale $\tau_d$ is
\begin{eqnarray}
\tau_d = \frac{1}{\sqrt{C}T}.
 \end{eqnarray}
 The heat capacity $C$ typically scales with the number $N$ of particles of the elementary detector and vanishes as $T \rightarrow 0$. Hence, the decay time scale $\tau_d$ decreases with $1/\sqrt{N}$ and increases  as $T \rightarrow 0$.

\subsection{Position measurements.} A typical record in a position measurement is the appearance of  a localized excitation in an underlying medium, such as a fluid or a lattice. For example,  wire chambers and semiconductor detectors are characterized by electric current excitations, bubble chambers by excitations in collective degrees of freedom. The position observable ${\pmb Q}$ corresponds to any variable that characterizes the location of the excitation. The size $\delta$ of the excitation determines the accuracy of the position sampling.

Degradation in a position measurement corresponds to the diffusion of the excitation in the underlying medium due to the random interaction between the medium's components. In a homogeneous system, the degradation function $\eta({\pmb Q}, s)$ should be the same for all points ${\pmb Q}$, so it does not depend on ${\pmb Q}$.

At high temperatures, diffusion in a homogeneous system is well described by the standard Brownian motion. The excitation is described by a probability density $\rho({\pmb q})$, where ${\pmb q}$ is the position coordinate in the underlying medium, and the probability density evolves with the propagator

  \begin{eqnarray}
  G({\pmb q}, 0; {\pmb q'}, s) = \frac{1}{(4\pi D |s|)^{3/2}} e^{- \frac{({\pmb q} - {\pmb q'})^2}{4D|s|}}, \label{diff}
  \end{eqnarray}
  where $D$ is the diffusion coefficient.

 A measurement record corresponds to a probability distribution $\rho_{\delta}({\pmb q})$ located around ${\pmb q} = 0$ with spread $\delta$. In a homogeneous medium the location of the measurement event does not affect the degradation, so there is no loss of generality in choosing ${\pmb q} = 0$. The corresponding persistence probability is
 \begin{eqnarray}
 |\eta(s)|^2 = \int d^3q d^3q'  G({\pmb q}, 0; {\pmb q'}, s) \rho_{\delta}({\pmb q})\rho_{\delta}({\pmb q}'). \label{etaposi}
 \end{eqnarray}
The persistence probability (\ref{etaposi}) can be estimated by substituting $({\pmb q} - {\pmb q'})^2$ with the spread $\delta^2$ in the propagator. Then,

\begin{eqnarray}
 |\eta(s)|^2 =  \frac{1}{(4\pi D |s|)^{3/2}} e^{- \frac{\delta^2}{4D|s|}}. \label{paposi}
 \end{eqnarray}

 Eq. (\ref{paposi}) implies that the decay time $\tau_d$ is of the order of $\delta^2/D$, i.e., it is inversely proportional to the diffusion coefficient $D$.

\section{Evaluation of the detection kernel}
In this section, we continue with the calculation of the detector kernel and we identify a simple expression for its general form. We use this expression, in order to simplify the probabilities associated to $n$ measurement records, and  to identify regimes that correspond to ideal measurements.

\subsection{Position localization}

Using the master equation approximation, the inner product
\begin{eqnarray}
W =  \langle \omega'| \hat{J}^{B}({\pmb q}') \sqrt{\hat{F}}({\pmb Q}, \mu)
e^{i\hat{h}s} \sqrt{\hat{F}}({\pmb Q}, \mu)  \hat{J}^{A}({\pmb q})|\omega'\rangle \nonumber \\
\end{eqnarray}
 that appears
in the detector kernel, Eq. (\ref{kernel2}), becomes
\begin{eqnarray}
W  = \eta(\mu, s) \langle \omega'| \hat{J}^{B}({\pmb q}') \hat{F}_1({\pmb Q}) \hat{F}_2( \mu)   \hat{J}^{A}({\pmb q})|\omega'\rangle, \label{WW}
\end{eqnarray}
where we employed  Eq. (\ref{ff1f2}). We assumed a homogeneous detector, so  the degradation function for a position measurement is ${\pmb Q}$-independent.

The currents $\hat{J}^A({\pmb q})$ define the interaction between the detector and the quantum field.  The vector $\hat{J}^A({\pmb q})|\omega'\rangle$ is typically characterized by  a microscopic length scale $\ell$ that corresponds to the range of the interaction. For example, $\ell$ may be the atomic scale (Bohr's radius) or  the nuclear scale. Whatever the value of $\ell$, it  is much smaller than the  {\em macroscopic} scale $\delta$ that characterizes  the localization of a position record. This means that
\begin{itemize}
\item[--] for $|{\pmb Q} - {\pmb q}| < \delta$, $\hat{F}_1({\pmb Q})  \hat{J}^{A}({\pmb q})|\omega'\rangle \simeq  \hat{J}^{A}({\pmb q})|\omega'\rangle$, and
\item[--]    for  $|{\pmb Q} - {\pmb q}| >>\delta$, $\hat{F}_1({\pmb Q})  \hat{J}^{A}({\pmb q})|\omega'\rangle \simeq  0$.
\end{itemize}

Thus, we can write
\begin{eqnarray}
\hat{F}_1({\pmb Q})  \hat{J}(q) |\omega\rangle = u_{\delta}({\pmb Q} - {\pmb q})\hat{J}(q) |\omega\rangle,
\end{eqnarray}
in terms of a sampling function $u_{\delta}: S \rightarrow {\pmb R}^+$ that satisfies  $\int d^3Q  u_{\delta}({\pmb Q}) = 1$.

Thus, the inner product (\ref{WW}) involves a product $\sqrt{u_{\delta}({\pmb Q} - {\pmb q}) u_{\delta}({\pmb Q} - {\pmb q'})}$. We choose  a Gaussian sampling function
\begin{eqnarray}
u_{\delta}({\pmb q}) = \frac{1}{(2\pi \delta^2)^{3/2}} e^{ - \frac{{\pmb q}^2}{2 \delta^2}}.
\end{eqnarray}
Then,
\begin{eqnarray}
\sqrt{u_{\delta}({\pmb Q} - {\pmb q}) u_{\delta}({\pmb Q} - {\pmb q'})} = u_{\delta} \left({\pmb Q} - \frac {{\pmb q} + {\pmb q'}}{2} \right) w_{\delta} ({\pmb q} - {\pmb q'}). \label{sqrtiden}
\end{eqnarray}
where
\begin{eqnarray}
w_{\delta}({\pmb q}) = e^{ - {\pmb q^2}/(8 \delta^2)}.
\end{eqnarray}

Assuming a scale of observation for ${\pmb Q}$ much larger than $\delta$ we can approximate $u_{\delta}({\pmb Q} - \bar{\pmb q})$   with a delta function $\delta^3({\pmb Q} - \bar{\pmb q})$---here $\bar{\pmb q} = ({\pmb q} +{\pmb q'})/2$ . Writing ${\pmb r} = {\pmb q} - {\pmb q'}$, the inner product (\ref{WW}) becomes
\begin{eqnarray}
W  = \eta(\mu, s)  \delta^3[{\pmb Q} - ({\pmb q} +{\pmb q'})/2] w_{\delta}({\pmb q}- {\pmb q}')  \langle \omega'| \hat{J}^{B}({\pmb q}')\hat{F}_2( \mu)   \hat{J}^{A}({\pmb q})|\omega'\rangle. \label{WW2}
\end{eqnarray}

\subsection{Matrix elements of observables}
To further simplify the inner product (\ref{WW2}), we   consider the Hilbert space ${\cal V}$ of test functions for the current operators $\hat{J}^A({\pmb q})$.   ${\cal V}$ is the space of square integrable functions $f_A: S \rightarrow {\pmb C}$, with an inner product
\begin{eqnarray}
(f|g) = \sum_A \int d^3q f_A^*({\pmb q})g_A({\pmb q}).
\end{eqnarray}
We employ a Dirac-like notation and we write the vectors of
 ${\cal V}$  as $| \cdot )$. The elements $|{\pmb q}, A)$ define  generalized eigenstates of the position operator on $S$.

 Then, we express the inner product $\langle \omega'|\hat{J}^B({\pmb q'}) \hat{F}_2(\mu)   \hat{J}^A({\pmb q})|\omega' \rangle$ on ${\cal K}$ as an inner product on ${\cal V}$,
\begin{eqnarray}
\langle \omega'|\hat{J}^B({\pmb q'}) \hat{F}_2(\mu)   \hat{J}^A({\pmb q})|\omega' \rangle  = C ( {\pmb q'},B|\hat{T}(\mu)|{\pmb q}, A),
\end{eqnarray}
where $C$ is a positive constant and $\hat{T}(\mu)$ are positive operators  on ${\cal V}$.
The constant $C$ is determined as follows.

We assume that all matrix elements of the operator $\hat{J}^A({\pmb q})$  between states of ${\cal K}_-$ vanish, because otherwise there would be no  meaningful correlations between microscopic particles and detection records.  Hence,

\begin{eqnarray}
(\hat{1} - \hat{E}) \hat{J}^A({\pmb q}) |\omega \rangle = 0
\end{eqnarray}
Since $\sum_{\mu}\hat{F}_2(\mu) = \hat{E}$,
\begin{eqnarray}
C \sum_{\mu} ( {\pmb q'}, B|\hat{T}(\mu)|{\pmb q}, A) = \langle \omega' | \hat{J}^B({\pmb q'}) \hat{J}^A({\pmb q})|\omega' \rangle. \label{ccor}
\end{eqnarray}

The correlation function is expected to vanish for $|{\pmb q} - {\pmb q}'| >> \ell$, where $\ell$ is the characteristic scale of the interaction. Hence, at scale much larger than $\ell$, it is proportional to a delta function

\begin{eqnarray}
\langle \omega | \hat{J}^B({\pmb q'}) \hat{J}^A({\pmb q})|\omega \rangle \simeq C \delta^3({\pmb q} - {\pmb q'}) \gamma^{AB} = C   ( {\pmb q'}, B|{\pmb q}, A)   , \label{cdef}
\end{eqnarray}
where $\gamma^{AB}({\pmb q})$ is a tensor that plays the role of the metric for the discrete degrees of freedom.
In most cases $\gamma^{AB}$ will be the Kronecker delta $\delta^{AB}$. However, when the collective index $A$  includes spacetime  or spinor indices,  $\gamma^{AB}$  may  be a non-trivial function of ${\pmb q}$.

Eq. (\ref{cdef}) serves as a definition for $C$. It also implies that the positive operators $\hat{T}(\mu)$ define a POVM on ${\cal V}$, since by Eq. (\ref{ccor}), $\sum_{\mu}  \hat{T}(\mu)= \hat{1}$.

Then, the inner product (\ref{WW2}) becomes
\begin{eqnarray}
W = C \eta(\mu, s)  \delta^3[{\pmb Q} - ({\pmb q} +{\pmb q'})/2] w_{\delta}({\pmb q}- {\pmb q}') ( {\pmb q'}, B|\hat{T}(\mu)|{\pmb q}, A). \label{WW3}
\end{eqnarray}

\subsection{General form of the detector kernel}

From Eqs. (\ref{kernel2}) and (\ref{WW3}), we obtain the following expression for the apparatus kernel,

\begin{eqnarray}
R(X, A; X', B| \tau, {\pmb Q}, \mu) = C \int ds g_{\sigma}(s) \eta(\mu, s) \int d^3r  w_{\delta}({\pmb r})
 \nonumber \\
\times  \delta^4[ X - {\cal E}(\tau +\frac{s}{2}, {\pmb Q} + \frac{\pmb r}{2} )]  \delta^4[ X' - {\cal E}(\tau -\frac{s}{2}, {\pmb Q} - \frac{\pmb r}{2} )]    ( {\pmb Q} - \frac{\pmb r}{2}, B|\hat{T}(\mu)|{\pmb Q} + \frac{\pmb r}{2}, A),\label{kernel5}
\end{eqnarray}
where we set ${\pmb r} - {\pmb q} - {\pmb q'}$.

When summing over all values of the observed variable $\mu$, a term of the form
\begin{eqnarray}
\sum_{\mu} ( {\pmb q'}, B|\hat{T}(\mu)|{\pmb q}, A) \eta(\mu,s)  \label{term}
\end{eqnarray}
appears in the kernel. This term  simplifies significantly when
 the degradation function $\eta(\mu, s)$ is $\mu$-independent. Then, the sum of Eq. (\ref{term}) equals $\delta^3({\pmb q'}- {\pmb q}) \gamma^{AB}({\pmb q}) \eta(s)$, and we obtain
\begin{eqnarray}
\sum_{\mu} R(X, A; X', B| \tau, {\pmb Q}, \mu) = \int ds C \gamma^{AB}   g_{\sigma}(s) \eta(s)
\delta^4[ X - {\cal E}(\tau +\frac{s}{2}, {\pmb Q} )]  \delta^4[ X' - {\cal E}(\tau -\frac{s}{2}, {\pmb Q}  )]
\end{eqnarray}

The parameter $C$ involves the coupling constants of the interaction between field and detectors and it is proportional to the total number of particles that left a detection record. When we normalize the probabilities by dividing with the fraction of the particles that have been detected $C$ drops out.

The detector kernel has a strong dependence on the temporal coarse-graining scale $\sigma$ and on the spatial coarse-graining scale $\delta$. A measuring event corresponds to a four-volume of order $\sigma \delta^3$ in $\Sigma \times {\pmb R}$. The corresponding spacetime four-volume depends on the detector's motion as encoded in the embedding function. We also recall that $\sigma >> \delta$, so that a single proper time parameter can be used for labeling the events in one detector.

To summarize, we found that  each measuring apparatus is characterized by
\begin{enumerate}[(i)]

\item its macroscopic motion in space-time encoded in the  embedding function ${\cal E}$,

\item a  POVM $\hat{T}(\mu)$ that is defined on the Hilbert space ${\cal V}$, describing the measurement  of all observables other than position,

\item  spatial resolution $\delta$ and    temporal resolution $\sigma$ of the spacetime coordinates, and

\item the degradation function $\eta(\mu, \tau)$.

\end{enumerate}

\section{Detection probabilities }

Next, we insert Eq. (\ref{kernel5}) for the apparatus kernel  into the probability densities (\ref{longh}), and identify idealized expression for the probabilities associated to $n$ measurement events of general validity.
 We first consider $n=1$  and then proceed to the most general case.
\subsection{A single measurement event}

We construct the probability density $P(\tau, {\pmb Q}, \mu)$  that   the detector records an event characterized by proper time $\tau$, spatial coordinate ${\pmb Q}$ and observable $\mu$,

\begin{eqnarray}
P(\tau, {\pmb Q}, \mu) = C \int d^3r ds g_{\sigma}(s) \eta(\mu, s) w_{\delta}({\pmb r}) G_{AB}[{\cal E}( \tau + \frac{s}{2}, {\pmb Q}+ \frac{\pmb r}{2});  {\cal E}(\tau - \frac{s}{2}, {\pmb Q}- \frac{\pmb r}{2})]
\nonumber \\
\times ( {\pmb Q} - \frac{\pmb r}{2}, B|\hat{T}(\mu)|{\pmb Q} + \frac{\pmb r}{2}, A), \label{1prob}
\end{eqnarray}

 By $G_{AB} (X, X')$ we denote the $2$-point function (\ref{2np})
\begin{eqnarray}
G_{AB} (X, X') := G(X,A| X', B) = Tr \left[ \hat{Y}_A(X) \hat{\rho}_0 \hat{Y}_B(X')\right].
\end{eqnarray}

For  observables $\mu$ such that $\eta(\mu, s)$ is $\mu$-independent, the integrated probability density is
 \begin{eqnarray}
 P(\tau, {\pmb Q}) :=  \sum_{\mu} P(\tau, {\pmb Q}, \mu) = C \int  ds g_{\sigma}(s) \eta(s) G[{\cal E}( \tau + \frac{s}{2}, {\pmb Q});  {\cal E}(\tau - \frac{s}{2}, {\pmb Q})], \label{summedprob}
 \end{eqnarray}
 where $G(X,X') = \gamma^{AB}G_{AB} (X, X') $. The same expression applies for the probability density  for measurements with  no   records other than position, i.e, for $\hat{F}_2(\mu) = \hat{E}$.

 The notion of an {\em ideal} measurement  is of particular significance in quantum measurement theory.  Ideal measurements are described  by probabilities that are insensitive to features of the measurement device. For example, the most general measurement of position is described in terms of a POVM that commutes with the position operator. In an ideal measurement, the POVM consists of the   spectral projectors of the position operator. In this case, it  does not depend on specific properties of the measuring apparatus, as for example, the localization scale of the measurement event.

The probabilities (\ref{1prob}) depend on several different parameters with different physical significance. Hence, there are several different types of ideal measurement.

\begin{enumerate}
\item {\em The point-like detector.}    In this regime,   the detector's world-tube collapses into a single time-like path, defined, say, by   ${\pmb Q} = 0$. Writing $X_0^{\mu}(\tau) = {\cal E}^{\mu}(\tau, 0)$, we obtain
    \begin{eqnarray}
 P(\tau, \mu)   = C \int ds g_{\sigma}(s) \eta(\mu, s) G_{AB}[X_0( \tau + \frac{s}{2}),  X_0( \tau - \frac{s}{2})] k^{AB}(\mu),
    \end{eqnarray}
    where the matrices $k^{AB}(\mu)$ are defined as
    \begin{eqnarray}
    k^{AB}(\mu) = \int d^3 r w_{\delta}({\pmb r}) ( - \frac{\pmb r}{2}, B|\hat{T}(\mu)| \frac{\pmb r}{2}, A).
 \end{eqnarray}
 In a point-like detector only {\em ultra-local observables} can be measured, i.e., observables that do not depend on derivative operators on the Hilbert space ${\cal V}$.
 An example of point-like detector model is the Unruh-Dewitt detector \cite{Unruh, Dewitt}.

\item {\em The fully coherent detector.} If the  characteristic scales of the functions $g_{\sigma}$, $w_{\delta}$ and $\eta$ are much larger than any
 microscopic scales appearing in the correlation function $G_{AB}$, but much smaller than the macroscopic scale of observation, we can   set $w_{\delta}\rightarrow 1 $,$g_{\sigma} \rightarrow 1$ and $\eta(\mu, s) = \eta(\mu, 0) = 1$, to obtain
    \begin{eqnarray}
    P(\tau, {\pmb Q}, \mu) = C\int ds  \int d^3r     G_{ΑΒ}[{\cal E}( \tau + \frac{s}{2}, {\pmb Q}+ \frac{\pmb r}{2});  {\cal E}(\tau - \frac{s}{2}, {\pmb Q}- \frac{\pmb r}{2})]
\nonumber \\
\times ( {\pmb Q} - \frac{\pmb r}{2}, B|\hat{T}(\mu)|{\pmb Q} + \frac{\pmb r}{2}, A). \label{cohdet}
    \end{eqnarray}

\item {\em The incoherent detector.} This condition  applies only to apparatuses that  record  no   observable $\mu$, so that Eq. (\ref{summedprob}) applies. In this regime,  the temporal coarse-graining parameter $\sigma$   is much smaller than any  temporal  parameter that characterizes the correlation function $G_{AB}$. This is possible when the characteristic decay time-scale in $\eta(s)$ is itself much smaller than any  temporal  parameter in $G_{AB}$.

Then, we substitute $g_{\sigma}$  by a delta function
     $g_{\sigma}(s) \rightarrow \sqrt{8\pi} \sigma \delta(s)$. Hence,

 \begin{eqnarray}
 P(\tau, {\pmb Q}, \mu)   = C'  G[{\cal E}({\pmb Q}, \tau),  {\cal E} ({\pmb Q}, \tau)]  \label{incohdet}
    \end{eqnarray}
where $C' = \sqrt{8\pi} \sigma    C $. The detection probability depends only on the diagonal elements of the correlation function. In Sec. 7, we  show that Glauber photo-detectors  \cite{Glauber} are special cases of incoherent detectors.

\end{enumerate}

\subsection{Time-integrated probabilities}

When the time instant of a measurement event is not recorded, we employ the time-integrated probability densities
\begin{eqnarray}
P({\pmb Q}, \mu) = \int_0^{\infty} d \tau P(\tau, {\pmb Q}, \mu).\label{timeintrange}
\end{eqnarray}
The range of time-integration in Eq. (\ref{timeintrange}) is $[0, \infty)$, since we assume that the initial preparation of the system corresponds to $\tau = 0$. Nonetheless, the probability density $P(\tau, {\pmb Q}, \mu)$ can be analytically extended to negative values of $\tau$. In some set-ups,
  $P(\tau, {\pmb Q}, \mu)$ is strongly suppressed for $\tau < 0$. This is the case, for example, in time-of-arrival measurements: particles are prepared in a spatially localized state with support on momenta directed towards the detector.   In such cases, we can extend the range of integration in Eq. (\ref{timeintrange})  to $(-\infty, \infty)$.

In two regimes,   the time-integrated probability Eq. (\ref{timeintrange}) takes a particularly simple form.
First, for the coherent detector, Eq. (\ref{cohdet}) and for $\tau$ integrated over $(-\infty, \infty)$,
\begin{eqnarray}
P({\pmb Q}, \mu) = C \int d^3r    Tr\left[\hat{V}_A(Q +\frac{{\pmb r}}{2}) \hat{\rho}_0 \hat{V}_B(Q-\frac{{\pmb r}}{2})\right]
\nonumber \\
\times ( {\pmb Q} - \frac{\pmb r}{2}, B|\hat{T}(\mu)|{\pmb Q} + \frac{\pmb r}{2}, A), \label{cohdet2}
\end{eqnarray}
where
\begin{eqnarray}
\hat{V}_A({\pmb Q}) = \int_{-\infty}^{\infty} d\tau \hat{Y}_A[{\cal E}(\tau, {\pmb Q})], \label{VAsum}
\end{eqnarray}
is the time-average of the composite operator $\hat{Y}_A$   along the detector's world volume.

Eq. (\ref{cohdet2}) is a consequence of
Eq. (\ref{longh}). However, the latter equation follows from an approximation by which we ignore a further classical smearing via convolution of the probability densities with respect to time, at a scale of $\sigma$ \cite{AnSav15a}. Ignoring this smearing may lead to problems in the definability if the integral over $\tau$ in Eq. (\ref{VAsum}) extends over the whole real axis. Without this approximation, the time-average  composite operator should be expressed as
 \begin{eqnarray}
\hat{V}_A({\pmb Q}) = \int_{-\infty}^{\infty} d\tau \sqrt{f_{\sigma}(\tau)} \hat{Y}_A[{\cal E}(\tau, {\pmb Q})], \label{VAsum2}
\end{eqnarray}
where $f_{\sigma}$ is a smearing function at the scale of $\sigma$.

The second regime where Eq. (\ref{timeintrange}) corresponds to the incoherent detector. By Eq. (\ref{incohdet}),
\begin{eqnarray}
P({\pmb Q}, \mu) = C' \gamma^{AB} \int_0^{\infty}d \tau Tr \left[ \hat{Y}_B({\cal E}(\tau, {\pmb Q})) \hat{\rho}_0 \hat{Y}_B({\cal E}(\tau, {\pmb Q}))\right] \label{incohdet2}
\end{eqnarray}

 Eqs. (\ref{cohdet2}) and (\ref{incohdet2}) apply to different physical situations. In Eq. (\ref{incohdet2}) all superpositions associated  to different instants of detection are decohered, while in Eq. (\ref{cohdet2}) they are preserved. Thus, the two equations may lead to different  predictions in physical configurations that are characterized by very small temporal parameters. This is the case for
   for particle oscillation experiments \cite{GePa, Ponte} in which time-evolution induces a very small time due to the interference of particle states with different mass. In Ref. \cite{AnSav12}, it was shown that the standard formula for particle oscillations requires the postulate of an incoherent detector as in
     Eq. (\ref{incohdet2}). In contrast, the consideration of  a coherent detector as in Eq. (\ref{cohdet2}) leads to a non-standard oscillation formula  \cite{AnSav10}.


\subsection{Multiple measurement events}

  Next, we evaluate the probability density (\ref{longh}) associated to $n$ detection events, each corresponding to a specific value of the proper time $\tau_i$, of the coordinate ${\pmb Q}_i$ and the observable $\mu_i$ of the $i$-th detector.   Each detector is characterized by an instantaneous space $S_i$, and
is spacetime motion is given by embedding function ${\cal E}_i: S_i \times {\pmb R} \rightarrow M$. The temporal coarse-graining scale of the $i$-th detector is denoted as $\sigma_i$, the spatial coarse-graining  scale  as $\delta_i$. and the associated the degradation function as $\eta_i(\tau_i)$.

 We find
\begin{eqnarray}
P(\tau_1, {\pmb Q}_1, \mu_1; \ldots; \tau_n, {\pmb Q}_n, \mu_n) = \int \prod_{i=1}^n (C_i  ds_i d^3r_i)
\nonumber \\
\prod_{i=1}^n \left[ w_{\delta_i}({\pmb r}_i)
g_{\sigma_i}(s_i) \eta_i(\mu_i, s_i)
( {\pmb Q}_i - \frac{{\pmb r}_i}{2}, B_i|\hat{T}(\mu_i)|{\pmb Q}_i + \frac{{\pmb r}_i}{2}, A_i)\right]
\nonumber \\
\times G
[{\cal E}_1(\tau_1 + \frac{s_1}{2}, {\pmb Q}_1+ \frac{{\pmb r}_1}{2}), A_1; \ldots ;{\cal E}_n( \tau_n + \frac{s_n}{2}, {\pmb Q}_n+ \frac{{\pmb r}_n}{2}), A_n| \nonumber \\
 {\cal E}_1(\tau_1 - \frac{s_1}{2}, {\pmb Q}_1 - \frac{{\pmb r}_1}{2}), B_1; \ldots ;
{\cal E}_n(\tau_n - \frac{s_n}{2}, {\pmb Q}_n+ \frac{{\pmb r}_n}{2}), B_n], \label{pqti}
\end{eqnarray}
where $G (X_1, A_1;\ldots ;X_n, A_n| X_1', B_1;\ldots ; X_n', B_n) $ is the $2n$-point function (\ref{2np}).

In absence of records $\mu_i$, or for records $\mu_i$ such that $\eta_i$ is $\mu_i$-independent, we obtain the integrated probabilities
\begin{eqnarray}
P(\tau_1, {\pmb Q}_1; \ldots; \tau_n, {\pmb Q}_n) : = \sum_{\mu_1,\ldots, \mu_n} P(\tau_1, {\pmb Q}_1, \mu_1; \ldots; \tau_n, {\pmb Q}_n, \mu_n) =
\int\prod_{i=1}^n  ds_i \prod_{i=1}^n \left[C_i
g_{\sigma_i}(s_i) \eta_i(s_i)\right] \nonumber \\
 G
[{\cal E}_1(\tau_1 + \frac{s_1}{2}, {\pmb Q}_1), A_1;\ldots; {\cal E}_n(\tau_n + \frac{s_n}{2}, {\pmb Q}_n), A_n|
 {\cal E}_1( \tau_1 - \frac{s_1}{2}, {\pmb Q}_1), A_1; \ldots ;
{\cal E}_n(\tau_n - \frac{s_n}{2}, {\pmb Q}_n), B_n], \label{pintn}
\end{eqnarray}

As in the case of the single measurement, we can identify different regimes for ideal measurements, namely, point-like, coherent and incoherent detectors. They follow from Eq. (\ref{pqti}) by an obvious application of the conditions presented in Sec. 3.3.1; their explicit expression is not given here.

\section{Particle detection  }
We consider the  probability density $P(\tau, {\pmb Q})$, Eq. (\ref{summedprob}) for the detection of relativistic particles of mass $m$, ignoring all other measurement records $\mu$. Thus, the only observables are the spacetime coordinates for a particle-detection event. This set-up generalizes the treatment of time-of-arrival measurements of Ref. \cite{AnSav12} by (i) considering different types of interaction and (ii) taking into account the detector's spacetime trajectories.

For simplicity, we assume spinless particles. The Hilbert space associated to a single particle is ${\cal H}_1 = L^2[{\pmb R}^3, d\mu(k)]$ where $d\mu(k) = \frac{d^3k}{2 (2\pi)^3 \omega_{\pmb k}}$ is the Lorentz invariant integration measure.  We consider a free scalar field $\hat{\phi}(X)$, so the field Hilbert space ${\cal F}$ is the bosonic Fock space associated to ${\cal H}_1$. The field $\hat{\phi}(X)$
  is defined   terms of the standard creation and annihilation operator on   ${\cal F}$:
\begin{eqnarray}
\hat{\phi}(X) = \int d\mu(k) \left(\hat{a}_k e^{i k \cdot X} + \hat{a}^{\dagger}_k e^{-ik\cdot X}\right),
\end{eqnarray}
where $k\cdot X = k_{\mu}X^{\mu}$, $k_{\mu} = (\omega_{\pmb k}, {\pmb k})$, $\omega_{\pmb k} = \sqrt{{\pmb k}^2+m^2}$. The creation and annihilation operators satisfy the commutation relations
\begin{eqnarray}
[\hat{a}_k, \hat{a}_{k'}] = [\hat{a}_k^{\dagger}, \hat{a}^{\dagger}_{k'}]  = 0 \hspace{1cm} [\hat{a}_k, \hat{a}^{\dagger}_{k'}] = (2\pi)^3 (2\omega_{\pmb k}) \delta^3({\pmb k} - {\pmb k}').
\end{eqnarray}

\subsection{Linear coupling}

First, we consider the case of a linear field-apparatus  coupling, that is, we identify the composite operator $\hat{Y}(X)$ with $\hat{\phi}(X)$. Since the interaction is linear to the creation and annihilation operators, this coupling corresponds to detection by particle absorption.

We obtain

\begin{eqnarray}
P(\tau, {\pmb Q}) = \int ds \eta(s) \langle \Psi| \hat{\phi}[{\cal E}( \tau- \frac{s}{2}, {\pmb Q})] \hat{\phi}[{\cal E}( \tau + \frac{s}{2}, {\pmb Q})]|\Psi\rangle, \label{pardet}
\end{eqnarray}
where we dropped the multiplicative constant $C$ and absorbed $g_{\sigma}$ into $\eta(s)$. In what follows, we assume that $\eta(s) = \eta(-s)$.

We consider an initial state with a definite number $N$ of particles. Then, Eq. (\ref{pardet}) becomes

\begin{eqnarray}
P(\tau, {\pmb Q}) = \int ds \eta(s) \Delta^+[{\cal E}(\tau + \frac{s}{2}, {\pmb Q}),  ({\cal E}( \tau - \frac{s}{2},{\pmb Q})] \nonumber \\
+ 2 N \int ds \eta(s)  \int d\mu(k) d\mu(k') \rho_1(k,k')
e^{-i k\cdot  {\cal E}( \tau + \frac{s}{2}, {\pmb Q}) + i k' \cdot {\cal E}(\tau + \frac{s}{2}, {\pmb Q})}
 \label{pardet2}
\end{eqnarray}
where
\begin{eqnarray}
\Delta^+(X,X') = \langle 0|\hat{\phi}(X') \hat{\phi}(X)|0\rangle = \int d \mu(k) e^{i k\cdot(X'-X)} \label{Wightman}
\end{eqnarray}
is the field's Wightman's function and
\begin{eqnarray}
\rho_1(k,k') = \frac{1}{N}\langle \Psi|\hat{a}^{\dagger}_{k} \hat{a}_{k'}|\Psi\rangle,
\end{eqnarray}
is the one-particle reduced density matrix associated to the state $|\Psi\rangle$. The reduced density matrix $\hat{\rho}_1$ is normalized as $\int d\mu(k) \rho_1(k,k) = 1$.

The first term in Eq. (\ref{pardet2}) is state-independent and describes particle creation from the vacuum as in the Unruh effect. It has been extensively studied in Ref. \cite{AnSav14}. This terms vanishes for motion along inertial trajectories, and it is suppressed exponentially with increasing mass $m$. In general, its contribution is very small, unless the detector's motion is characterized by time-scales much smaller than the coarse-graining time scale $\sigma$. We will ignore this term in what follows.

The second term in Eq. (\ref{pardet2}) depends only on the one-particle reduced density matrix.
  This means that  no correlations between particles can be accessed from a single local measurement.

\subsection{The ideal time-of-arrival distribution}

 We consider a static detector located far away from a particle source, so that only particles with momentum along the source-detector axis can be recorded. Then, we can  restrict   to a single spatial dimension. The corresponding embedding function   is $({\cal E}^0, {\cal E}^1) = (\tau, Q)$, where $Q > 0$ denotes the distance of the detector from the particle source. Then, Eq. (\ref{pardet2}) becomes
\begin{eqnarray}
P(\tau, Q) =    2N \int d \mu(k) d\mu(k') \rho_1(k',k) \tilde{\eta}\left(\frac{\omega_k + \omega_{k'}}{2}\right) e^{i(\omega_k - \omega_{k'}) \tau + i (k'-k) Q }, \label{toa}
\end{eqnarray}
where $\tilde{\eta}$ is the Fourier transform of $\eta(s)$. Since we work in one spatial dimension, $d\mu(k) = \frac{dk}{(2\pi)(2 \omega_k)}$.

We consider the time-integrated probability density $P(Q)= \int_0^{\infty}  d \tau P(\tau, Q)$. For an initial state with support only on positive values of $k$, and localized at $x < Q$, we can extend the range of integration for $\tau$ to $(-\infty, \infty)$. Then
\begin{eqnarray}
P(Q) =  N \int_0^{\infty} d \mu(k)  \rho_1(k, k) \frac{\tilde{\eta}(\omega_k)}{|k|} \label{PQ}
\end{eqnarray}

Eq. (\ref{PQ}) implies that $\tilde{\eta}(\omega_k)/|k|$ is proportional to the absorption coefficient $\alpha(\omega)$ in the rest frame of the detector. The absorption coefficient is defined as the fraction of particles with energy $\omega$ absorbed per unit length at the detector. Hence,
\begin{eqnarray}
\tilde{\eta}(\omega) = A \alpha(\omega) \sqrt{\omega^2 - m^2}, \label{etaa}
\end{eqnarray}
for some constant $A$. Eq. (\ref{etaa}) imply that the degradation function for a detector can be reconstructed from the  measurement of its absorption coefficient, i.e., from information that is macroscopically accessible.

An idealized limit for this measurement corresponds to a constant absorption coefficient. We  also consider a constant value of $Q$ and normalize the probability distribution $P(\tau, Q)$ as $\int_{-\infty}^{\infty} d \tau P(\tau, Q) = 1$. Then,
\begin{eqnarray}
P(\tau, Q) = \int d \mu(k) d\mu(k') \rho_1(k, k') \sqrt{(\omega_k + \omega_{k'})^2 -4m^2} e^{-i(  \omega_{k'} - \omega_k) \tau + i (k'-k) Q }.  \label{toa2}
\end{eqnarray}

We consider an initial pure state $\rho_1(k, k') = \psi_0(k) \psi_0^*(k')$ that is well localized in momentum,  so that we can approximate $(\omega_k+\omega_{k'})^2-4m^2 \simeq 4 |k||k'|$. Then, Eq. (\ref{toa2}) becomes
\begin{eqnarray}
P(\tau, Q) = \left| \int \frac{\sqrt{k} dk}{\omega_k} \psi_0(k)  e^{-i  \omega_{k}  \tau + i k Q }\right|^2.  \label{toa3}
\end{eqnarray}

Eq. (\ref{toa3}) is the POVM derived in Ref. \cite{AnSav12}. It  generalizes  Kijowski 's POVM for the time of arrival \cite{Kij74} to relativistic particles.


\subsection{Moving detector}

For a general detector trajectory, we expand the embedding function ${\cal E}^{\mu}({\pmb Q}, \tau+s)$ around $\tau$ and keep terms to first order in $s$,
\begin{eqnarray}
{\cal E}^{\mu}({\pmb Q}, \tau+s) = {\cal E}^{\mu}({\pmb Q}, \tau) + u^{\mu}(\tau) s ,  \label{1stord}
\end{eqnarray}
where $u^{\mu} = \dot{{\cal E}}^{\mu}$. Eq. (\ref{1stord}) is a good approximation if $a \sigma << 1$, where $a$ is the proper acceleration of the path.

 Then, Eq. (\ref{pardet2}) becomes

\begin{eqnarray}
P(\tau, {\pmb Q}) = 2 N  \int d\mu(k) d\mu(k') \rho_1(k',k) \tilde{\eta} \left( \frac{(k+k') \cdot u(\tau)}{2} \right)  e^{i (k'-k) \cdot  {\cal E}(\tau, {\pmb Q})}, \label{ptq4}
\end{eqnarray}
i.e., $\tilde{\eta}$ is evaluated for energies in the rest frame of the detector.
In the regime of the ideal detector of Eq. (\ref{toa2}), Eq. (\ref{ptq4}) becomes
\begin{eqnarray}
P(\tau, {\pmb Q}) = \int d\mu(k) d\mu(k') \rho_1(k',k)  \sqrt{[(k+k')\cdot u(\tau)]^2 - 4m^2   } e^{i (k'-k) \cdot  {\cal E}(\tau, {\pmb Q})}, \label{ptq5}
\end{eqnarray}
Eq. (\ref{ptq5}) defines the ideal probability distribution for the time-of-arrival in a moving detector.

\subsection{Quadratic coupling}
Next, we consider the case of quadratic field-apparatus  coupling, that is, we identify the composite operator $\hat{Y}(X)$ with the normal-ordered field product $:\hat{\phi}^2(X):$. The relevant interaction term involves one creation and one annihilation operator of the field and it corresponds to detection through particle scattering.

We consider a single-particle initial state for the field
\begin{eqnarray}
|\Psi\rangle = \int d\mu(k) \psi_0(k) \hat{a}^{\dagger}_k|0\rangle.
\end{eqnarray}
Then, Eq. (\ref{summedprob}) yields
\begin{eqnarray}
P(\tau, {\pmb Q}) = 2 \int ds \eta(s) \left(\Delta^+[{\cal E}(\tau + \frac{s}{2}, {\pmb Q}),  ({\cal E}( \tau - \frac{s}{2},{\pmb Q})]\right)^2 \nonumber \\
+ 8 \int ds \eta(s)  \int d\mu(k) d\mu(k') \psi_0(k)^* \psi_0(k')\Delta^+[{\cal E}(\tau + \frac{s}{2}, {\pmb Q}),  ({\cal E}( \tau - \frac{s}{2},{\pmb Q})]
e^{-i k\cdot  {\cal E}( \tau + \frac{s}{2}, {\pmb Q}) + i k' \cdot {\cal E}(\tau + \frac{s}{2}, {\pmb Q})}.
 \label{pardet3}
\end{eqnarray}
The first term in Eq. (\ref{pardet3}) is state-independent and describes particle creation from the vacuum as in the Unruh effect.  This terms  is suppressed exponentially with increasing mass $m$. In general, its contribution is very small, unless the detector's motion is characterized by time-scales of the order of  the coarse-graining time scale $\sigma$ or smaller. We shall ignore it in what follows.

For a detector on an  inertial motion, the Wightman function Eq. (\ref{pardet3}) becomes
\begin{eqnarray}
\Delta^+(s) = \frac{1}{4\pi^2} \int_0^{\infty} \frac{k^2 dk}{\omega_k} e^{-i\omega s}
\end{eqnarray}

We define $\kappa(s) := 4 \eta(s) \Delta^+(s)$. Then, we  restrict to one spatial dimension, with an embedding  function   $({\cal E}^0, {\cal E}^1) = (\tau, Q)$. Then, Eq. (\ref{pardet2}) becomes
\begin{eqnarray}
P(\tau, Q) =    2 \int d \mu(k) d\mu(k') \rho_1(k',k) \tilde{\kappa}\left(\frac{\omega_k + \omega_{k'}}{2}\right) e^{i(\omega_k - \omega_{k'}) \tau + i (k'-k) Q }, \label{toa5}
\end{eqnarray}
where $\tilde{\kappa}$ is the Fourier transform of $\kappa$.

Eq. (\ref{toa5}) coincides with Eq. (\ref{toa}) modulo the identification of $\tilde{\kappa}$ with $N \tilde{\eta}$. This implies that the quadratic field-detector coupling also leads to the
 ideal distribution (\ref{toa2}).

\section{Photo-detection}
Next, we apply our formalism to the case of photo-detection.  The measured system is   the quantum electromagnetic field
\begin{eqnarray}
\hat{A}_{\mu}(X) = \int d\mu(k) \sum_{r=1}^2 \left[ \hat{a}_{k,r} \epsilon_{\mu}(k,r) e^{i k\cdot X} + \hat{a}^{\dagger}_{k,r} \epsilon_{\mu}(k, r) e^{-i k\cdot X} \right],
\end{eqnarray}
expressed in terms of the standard creation and annihilation operators;
$\epsilon^{\mu}(k, 1)$  and $\epsilon^{\mu}(k, 2)$  are  two orthogonal and transverse real-valued polarization vectors.

\subsection{Detection probabilities}

We consider an interaction between the electromagnetic field and the measuring apparatus of the form
\begin{eqnarray}
\hat{V} = \int d^3 x \hat{F}_{\mu \nu} ({\pmb x}) \otimes \hat{M}^{\mu \nu}({\pmb x}),
\end{eqnarray}
 where $\hat{F}_{\mu \nu} = \partial_{\mu} \hat{A}_{\nu} - \partial_{\nu} \hat{A}_{\mu}$ is the field-strength tensor, and $\hat{M}^{\mu \nu}$ is the relativistic magnetization-polarization tensor associated to the measuring apparatus.

 The electric polarization is typically much larger than the magnetization in the rest frame of the detector. Thus, the dominant contribution to the field-detector coupling corresponds to a current  $\hat{ \mathfrak{ p}}^i({\pmb q})$  that represents polarization density in the rest frame of the detector. In this case, the stationarity condition, Eq. (\ref{jcu}) takes the form
\begin{eqnarray}
\hat{M}^{\mu \nu}({\pmb x}) e^{-i \hat{h} \tau} |\omega\rangle = \int d^3q \left(\dot{\cal E}^{\mu} {\cal E}^{\nu}_i - \dot{\cal E}^{\nu} {\cal E}^{\mu}_i \right) \hat{\mathfrak{p}}^i({\pmb q}) \delta^3(x^i - {\cal E}^i ({\pmb q}, t))|\omega'\rangle,
\end{eqnarray}
for some state vector $|\omega'\rangle$.

Following the reasoning of Secs. 3 and 4, we derive the (unnormalized) probability density
  $P( \tau, {\pmb Q})$ of detection at time $\tau$ and position  ${\pmb Q}$,
\begin{eqnarray}
P( \tau, {\pmb Q}) = \int ds g_{\sigma}(s)  \eta(s) G[{\cal E}({\pmb Q}, \tau+ \frac{s}{2}), {\cal E}({\pmb Q}, \tau- \frac{s}{2})]. \label{pqtem}
\end{eqnarray}
The correlation function
\begin{eqnarray}
G(X,X') = h^{ij}   Tr \left[ \hat{E}_i(X') \hat{\rho}_0 \hat{E}_j(X) \right],
\end{eqnarray}
is defined in terms of the electric field vector in the detector's rest frame.
\begin{eqnarray}
\hat{E}_i = \frac{1}{2} \left(\dot{\cal E}^{\mu} {\cal E}^{\nu}_i - \dot{\cal E}^{\nu} {\cal E}^{\mu}_i \right)\hat{F}_{\mu \nu}
\end{eqnarray}
the inverse $h^{ij}$ of the induced metric $h_{ij} = {\cal E}^{\mu}_i {\cal E}^{\nu}_k\eta_{\mu\nu}$ on the instantaneous space $S$, and the initial state of the field $\hat{\rho}_0$.

Eq. (\ref{pqtem}) applies to any photo-detector, irrespective of its state of motion. It is important to note that the electric field depends on ${\pmb Q}$ and $\tau$ not only through its argument, but also through the tensor  $\frac{1}{2} \left(\dot{\cal E}^{\mu} {\cal E}^{\nu}_i - \dot{\cal E}^{\nu} {\cal E}^{\mu}_i \right)$ that projects onto the instantaneous space.

\subsection{Static detectors}

Next, we specialize to  static detectors, characterized by an embedding function ${\cal E} (\tau, {\pmb Q}) = (\tau, {\pmb Q})$. In this case, the tensor $\dot{\cal E}^{\mu} {\cal E}^{\nu}_i - \dot{\cal E}^{\nu} {\cal E}^{\mu}_i$ does not depend on ${\pmb Q}$ and $\tau$, and $h_{ij} = \delta_{ij}$ . The electric field  becomes
\begin{eqnarray}
\hat{ \pmb E}(X) = i \int d\mu(k) \omega_k \sum_{r=1}^2 \left[ \hat{a}_{k,r} {\pmb f}(k,r) e^{i k\cdot X} - \hat{a}^{\dagger}_{k,r} {\pmb  f} (k,r) e^{-i k\cdot X} \right],
\end{eqnarray}
where $\omega_k = \dot{\cal E}^{\mu} k_{\mu} = |{\pmb k}|$ and
\begin{eqnarray}
f_i(k,r) = (\dot{\cal E}^{\mu} {\cal E}^{\nu}_i - \dot{\cal E}^{\nu} {\cal E}^{\mu}_i) k_{\mu} \epsilon_{\nu}(k,r),
\end{eqnarray}
are the  unit  3-vectors  that describe polarization.

It is convenient to proceed by introducing the coherent states  $|z\rangle$  of the electromagnetic field. They are defined as eigenvalues of the annihilation operator

\begin{eqnarray}
\hat{a}_{k,r}|z\rangle = z_r(k)|z\rangle,
\end{eqnarray}
in terms of a square-integrable complex-valued function $z_r(k)$.

We   employ the $P$ representation for the initial state
\begin{eqnarray}
\hat{\rho} = \int Dz f_P(z) |z\rangle \langle z|, \label{psymbol}
\end{eqnarray}
 where $f_P(z)$ is a functional known as the $P$-symbol of  quantum state $\hat{\rho}$ \cite{Kla85}. The functional integration in Eq. (\ref{psymbol}) is defined in terms of the Gaussian  measure that defines the Bargmann representation of the quantum field \cite{berezin}.

The   probability density Eq. (\ref{pqtem}) is expressed as
\begin{eqnarray}
P(\tau, {\pmb Q}) = \int Dz f_P(z) P_z( \tau, {\pmb Q}),
\end{eqnarray}
where $P_z( \tau, {\pmb Q} $ is the probability density Eq. (\ref{pqtem}) evaluated for a coherent state. Substituting in Eq. (\ref{pqtem}),  we find that $P_z$ is a sum of three contributions
\begin{eqnarray}
P_z( \tau, {\pmb Q}) = P_z^{(0)}( \tau, {\pmb Q}) + P_z^{(1)}( \tau, {\pmb Q}) + P_z^{(2)}( \tau, {\pmb Q}), \label{probphoto}
\end{eqnarray}
where
\begin{eqnarray}
P_z^{(0)}( \tau, {\pmb Q}) &=& \int d \mu(k) \tilde{\eta}(\omega_k)
\nonumber \\
P_z^{(1)}( \tau, {\pmb Q}) &=& 2   \int d\mu(k) d\mu(k')      \omega_k \omega_{k'}  \tilde{\eta}\left(\frac{\omega_k +\omega_{k'}}{2}\right)
{\pmb \zeta}^*_{k'} \cdot {\pmb \zeta}_k
e^{i({\pmb k}- {\pmb k}')\cdot {\pmb Q} - i (\omega_k-\omega_{k'})\tau} \nonumber \\
P_z^{(2)}( \tau, {\pmb Q}) &=& - 2  \int   d\mu(k) d\mu(k')   \omega_k \omega_{k'}  \tilde{\eta}\left(\frac{\omega_k - \omega_{k'}}{2}\right) \mbox{Re} \left[ {\pmb \zeta}_k \cdot {\pmb \zeta}_{k'} e^{i({\pmb k}+ {\pmb k}')\cdot {\pmb Q} - i (\omega_k+\omega_{k'})\tau }\right],
\end{eqnarray}
We wrote ${\pmb \zeta}_k =\sum_{r=1}^2 z_r(k) {\pmb f}(k,r)$,  and we  absorbed the $g_{\sigma}$ term into the degradation function $\eta(s)$.

The meaning of the three terms in
  Eq. (\ref{probphoto}) is the following.

  \begin{itemize}

\item   The term $P_z^{(0)}$ is state-independent.  It corresponds to a background noise of `false alarms' in the detector
    \cite{PeTe} and its contribution to the detection probability is negligible.

  \item The  term $P_z^{(1)}$ is generated by the terms in the density matrix that corresponds to a fixed number of photons, i.e., by the restriction of the density matrix in the $N$ photon subspaces. It has the same form with the time-of-arrival probability distribution (\ref{toa}) for $m = 0$.

  \item The  term  $P_z^{(2)}$ corresponds  to  superpositions   in the number of photons. It will be denoted as $P_2( \tau, {\pmb Q})$.

  \end{itemize}

\subsection{The Rotating Wave Approximation and the Glauber detector}

The second term in  Eq. (\ref{probphoto}) involves a double momentum integral over $e^{-i(\omega_k - \omega_{k'})\tau}$ (co-rotating terms), while the  third term involves a double integral over $e^{-i(\omega_k + \omega_{k'})\tau}$ (counter-rotating terms). The dominant contributions to the double integral arise from values of $k$ and $k'$ that correspond to a slowly varying phase. This implies that co-rotating terms typically dominate over counter-rotating ones.  It is therefore meaningful to invoke the   Rotating Wave Approximation (RWA) and drop the contribution of the counter rotating terms.

The RWA is a defining approximation of Glauber's theory of photo-detection. The domain of validity of the RWA in photodetection has been  a matter of some debate. Some authors have shown that   photodetection probabilities that are obtained with  the RWA  appear to violate causality at short times and they proposed
corrections to Glauber's  theory \cite{ByTa}. Others indicate that a different form of the RWA in photodetection theory can guarantee causality \cite{MJF95}.  A fundamental problem of the RWA  is that it corresponds to a Hamiltonian that has no lower bound, and thus it is thermodynamically unacceptable \cite{FoCo}. However,  it has been suggested that the proper renormalization of the interaction between detector and quantum field may lead to an effective RWA dynamics \cite{CoPi}. Regarding the degree of accuracy of the RWA and related approximations, see, Ref. \cite{FCAH10} and references therein.

To examine the validity of the RWA, we consider a coherent state that corresponds to a pulse of mean wave-number ${\pmb k}_0$ localized around ${\pmb x} = 0$ at $t = 0$.  We choose a Gaussian
\begin{eqnarray}
{\pmb \zeta}_k = (2\pi)^3 {\pmb \zeta}_0 (2 \pi \Delta^2)^{-3/2} e^{ - \frac{ ({\pmb k} - {\pmb k}_0)^2}{2 \Delta^2}},
\end{eqnarray}
where ${\pmb \zeta}_0$ is a constant complex valued vector and $\Delta$ is the spectral width of the pulse.

First, we evaluate the term $P_z^{(1)}(\tau, {\pmb Q})$ in Eq. (\ref{probphoto}). We employ a saddle point approximation. We expand $\omega_k = \omega_{k_0} + {\pmb v}_{k_0} \cdot ({\pmb k} - {\pmb k}_0)$, where  ${\pmb v}_k = {\pmb k}/|{\pmb k}|$ is the three- velocity vector for a photon. Then, we assume  that $\tilde{\eta}$ varies slowly with $\omega$, so that the contribution of the degradation function is a constant $\tilde{eta}(\omega_{k_0})$, we find
\begin{eqnarray}
P_z^{(1)}(\tau, {\pmb Q}) = \frac{1}{2} |{\pmb \zeta}_0|^2 \tilde{\eta}(\omega_{k_0}) e^{- \Delta^2 ({\pmb Q} - {\pmb v}_{k_0} \tau)^2},
\end{eqnarray}
where ${\pmb v}_k = {\pmb k}/|{\pmb k}|$ is the three- velocity vector for a photon.

We also evaluate the term $P_2(\tau, {\pmb Q}) $ in the saddle-point approximation. We find
\begin{eqnarray}
P_z^{(2)}(\tau, {\pmb Q}) = - \frac{1}{2} e^{ - \frac{ ({\pmb k} - {\pmb k}_0)^2}{2 \Delta^2}}  \mbox{Re} \left[ {\pmb \zeta}_0^2 e^{i {\pmb k}_0\cdot {\pmb Q} - i \omega_{k_0}\tau }\right] \left( \int_{-\infty}^{\infty} ds \eta(s) e^{-\Delta^2 s^2}\right). \label{p2rwa}
\end{eqnarray}

The key point is that the oscillatory terms in Eq. (\ref{p2rwa}) must be averaged over a spatial region of size $\delta^3$ and for times of order $\sigma >> \delta$. For Gaussian smearing functions, this averaging leads to  suppression factors of order $e^{-\delta^2 |{\pmb k}|_0^2} e^{-\sigma^2 \omega_{k_0}^2}$. Thus, the contribution of the term $P_z^{(2)}$ to the probability density (\ref{probphoto}) is suppressed, and the RWA  is justified. We expect that the suppression of the counter-rotating terms due to spatial and temporal coarse-graining in the detector holds for a large class of initial states.

 The RWA is equivalent with the rewriting of  Eq. (\ref{pqtem}) as
\begin{eqnarray}
P( \tau, {\pmb Q}) = \int ds    \eta(s) G_{RWA}[{\cal E}(\tau+ \frac{s}{2}, {\pmb Q}), {\cal E}( \tau- \frac{s}{2}, {\pmb Q} )] . \label{pqtem2}
\end{eqnarray}
Eq. (\ref{pqtem2}) involves a different correlation function
\begin{eqnarray}
G_{RWA}(X, X') = 2 \delta^{ij} \langle \Psi|\hat{E}_i^{(-)}(X') \hat{E}_j^{(+)}(X) |\Psi\rangle,
\end{eqnarray}
where $\hat{E}_i^{(+)}$ and $\hat{E}_i^{(-)}$ denote the positive and negative frequency part of $\hat{E}_i$ respectively.

For an incoherent detector,  the probability distribution $P( \tau, {\pmb Q})$ becomes
\begin{eqnarray}
P( \tau, {\pmb Q}) = G_{RWA}[{\cal E}(\tau, {\pmb Q}), {\cal E}( \tau, {\pmb Q} )],
\end{eqnarray}
and, thus, coincides with the  standard formula of Glauber's photo-detection theory.

For a single measurement event we recover Glauber's photo-detection theory by assuming (i) the RWA   and (ii) the incoherent detector limit. As we showed, the RWA is not an independent assumption, as it arises due to the spatial and temporal coarse-graining of the detector for a large class of initial states. Thus, the defining assumption of Glauber's theory is that the temporal coarse-graining scale is much smaller than any time parameter characterizing the initial state of the field.

 In Glauber's theory, the $n$-time probabilities
 \begin{eqnarray}
P(\tau_1, {\pmb Q}_1, \mu_1; \ldots; \tau_n, {\pmb Q}_n, \mu_n) = G_{RWA}[{\cal E}(\tau_1, {\pmb Q}_1), \ldots, {\cal E}(\tau_n, {\pmb Q}_n); {\cal E}( \tau_1, {\pmb Q}_1), \ldots , {\cal E}( \tau_n, {\pmb Q}_n)] \label{glaubern}
\end{eqnarray}
 are defined in terms of correlation functions
 \begin{eqnarray}
 G_{RWA}(X_1, \ldots, X_n|X_1', \ldots, X_n') = \nonumber \\ \delta^{i_1j_1} \ldots \delta^{i_nj_n} Tr \left(\hat{E}_{i_n}^{(+)}(X_n)  \ldots \hat{E}_{i_1}^{(+)}(X_1) \hat{\rho}_0 \hat{E}_{j_1}^{(-)}(X'_1) \ldots \hat{E}_{j_n}^{(-)}(X'_n)\right).
 \end{eqnarray}

Eq.  (\ref{glaubern}) follows from Eq. (\ref{pintn}) assuming the RWA and the incoherent detector limits. However, we are skeptical about the validity of the RWA for general set-ups of $n$ measurements. The RWA may  misrepresent the effects of retarded propagation of the electromagnetic field \cite{DPP06}, hence, it may fail in configurations that involve several
 several detection events that are spacelike separated.  In contrast,  Eq. (\ref{pintn}) is applicable in such systems.

\section{Spin measurements}
Our last example involves the calculation of probabilities for spin measurements in relativistic electrons. This issue is of particular interest for relativistic quantum information because there are different proposals about the operator representation of spin measurements in relativistic systems \cite{spinop, spinop2}.

Spin measurements are often described in terms of the  Pauli-Lubanski vector  \cite{PaLu} and related operator. This description is covariant, in the sense that the expectation values for the spin measurements are the same in all reference frames. However, as noted in Ref. \cite{SaVe12}, there is no known physical coupling of relativistic particles  to the detector compatible with this description. In fact, a careful description of spin measurements (as, for example, in Stern-Gerlach-type experiments \cite{PTW13})   is important in order to avoid violations of relativistic causality \cite{SaVe13}.

Our starting point in this section is the QFT description of  particles of mass $m$ and spin $s = \frac{1}{2}$.
We consider the free Dirac field,
\begin{eqnarray}
\hat{\psi}(X) = \hat{\psi}^{(+)}(X) +  \hat{\psi}^{(-)}(X), \hspace{1.2cm} \hat{\bar{\psi}}(X) = \hat{\bar{\psi}}^{(+)}(X) +  \hat{\bar{\psi}}^{(-)}(X) \\
\hat{\psi}^{(+)}(X) = \sum_{r=1}^{2} \int d\mu(k) \hat{c}_r(k) u(k,r) e^{ik\cdot X} \hspace{1cm} \hat{\psi}^{(-)}(X) =   \sum_{r=1}^{2} \int d\mu(k) \hat{d}^{\dagger}_r(k) \bar{v}(k,r) e^{-ik\cdot X} \\
\hat{\bar{\psi}}^{(+)}(X) = \sum_{r=1}^{2} \int d\mu(k) \hat{d}_r(k) v(k,r) e^{ik\cdot X} \hspace{1cm} \hat{\bar{\psi}}^{(-)}(X) =   \sum_{r=1}^{2} \int d\mu(k) \hat{c}^{\dagger}_r(k) \bar{u}(k,r) e^{-ik\cdot X}
\end{eqnarray}
where $\hat{c}_r(k)$ and $\hat{d}_r(k)$ are annihilation operators for electrons and positrons respectively. The four-vectors $k$  satisfy $k^2 =- m^2$. The Dirac spinors $u(k,r)$ and $v(k,r)$ are normalized as
\begin{eqnarray}
\bar{u}(k,r) u(k, r) = 2 m \delta_{rr'} \hspace{1cm}  \bar{u}^{\dagger}(k,r) u(k, r) = 2 \omega_k \delta_{rr'},  \label{unorm}
\\
\bar{v}(k,r) v(k, r) = -2 m \delta_{rr'} \hspace{1cm}  \bar{v}^{\dagger}(k,r) v(k, r) = 2 \omega_k \delta_{rr'},
\end{eqnarray}
where $\omega_k = |{\pmb k}|$.

We consider an interaction between the electron field and the detector of the form
\begin{eqnarray}
\hat{V} = \int d^3x \left( \hat{\bar{\chi}}(x) \otimes \hat{\psi}(x) + \hat{\chi}(x) \otimes \hat{\bar{\psi}}(x) \right),
\end{eqnarray}
where $\hat{\chi} (x), \hat{\bar{\chi}}(x)$ are currents defined on the detector Hilbert space ${\cal K}$.

The stationarity condition, Eq. (\ref{jcu}), takes the form

\begin{eqnarray}
\hat{\bar{\chi}}({\pmb x}) e^{-i \hat{h} \tau} |\omega\rangle = \int d^3q \hat{\bar{\chi}}({\pmb q}) \delta^3(x^i - {\cal E}^i ({\pmb q}, t))|\omega'\rangle, \\ \label{jcu2}
\hat{\chi}({\pmb x}) e^{-i \hat{h} \tau} |\omega\rangle = \int d^3q \hat{\chi}({\pmb q}) \delta^3(x^i - {\cal E}^i ({\pmb q}, t))|\omega'\rangle,
\end{eqnarray}
for some $|\omega' \rangle \in {\cal K}$ and operators $\hat{\eta}({\pmb q})$ and $\hat{\bar{\chi}}({\pmb q})$ ${\cal K}$.

We assume that the detection signal is solely due to the absorption of electrons and not due to the absorption of positrons.  This implies that
 the operator $\hat{\chi}({\pmb q})$ that couples to the annihilation operator for positrons annihilates the initial state $|\omega'\rangle$ of the detector.

We consider an initial state $\hat{\rho}_0$ for the field that contains   $N$  electrons and zero positrons. Then, the only correlation functions that contributes into the one-event probability density Eq. (\ref{1prob}) are of the form $Tr \left(\hat{\psi}^{(+)}(X)\hat{\rho}_0 \hat{\bar{\psi}}^{(-)}(X')\right)$. Hence, all relevant information about the initial state is contained in the one-electron reduced density matrix
\begin{eqnarray}
\hat{\rho}_1(k,r;k',r') = \frac{1}{N} Tr \left[\hat{c}_r(k) \hat{\rho}_0 \hat{c}^{\dagger}_{r'}(k') \right] ,
\end{eqnarray}
which is normalized as
\begin{eqnarray}
\sum_r \int d\mu(k) \hat{\rho}_1(k,r; k,r)  = 1.
\end{eqnarray}

Next, we identify the Hilbert space ${\cal V}$ of test-functions for the current operators that was introduced in Sec. 4.2.  ${\cal V}$ consists of test functions $f_A$ on the instantaneous space $S$, where $A$ is a four-spinor index. A general function $f_A$ can be analyzed in terms of the Dirac spinors $u(k,r)$ and $v(k,r)$ as
\begin{eqnarray}
f_A({\pmb q}) =    \sum_{r=1}^2 \int d \mu(k)    \left( e^{i{\pmb k} \cdot {\pmb q}} u_A(k,r) \phi_r(k) + e^{i{\pmb k} \cdot {\pmb q}} v_A(k,r) \chi^*_r(k)\right),
\end{eqnarray}
where $\phi_r(k)$ and $\chi_r(k)$ are elements of the Hilbert space ${\cal H}_{1/2}$ that carries the irreducible representation of the Poincar\'e group for $s = \frac{1}{2}$.

 It is important to note that the positive operators $\hat{T}(\mu)$ are defined on the Hilbert space ${\cal V} = {\cal H}_{1/2} \oplus  \bar{\cal H}_{1/2}$ rather than on ${\cal H}_{1/2}$ as one would expect by a particle treatment of the electron, without taking its QFT coupling into account.   Since a  spin  measurement can only have two outcomes, the POVM $\hat{T}(\mu)$  can always be expressed as
  \begin{eqnarray}
  \hat{T}(\mu) = \frac{1}{2} \left( \hat{1} + \mu \hat{\Sigma}\right), \label{tsigma}
  \end{eqnarray}
  where $\mu = \pm 1$ and $\hat{\Sigma}$ is an operator on ${\cal V}$. The specification of $\hat{\Sigma}$ and its covariance properties depend on the detailed model for the spin measurement, and will be considered in a different publication.




Next, we specialize to the case of a static detector and   we express the embedding function ${\cal E}(\tau, {\pmb Q}) = (\tau, {\pmb Q}) $. We consider a $\mu$-independent degradation function $\eta$, and we evaluate the probability   density (\ref{1prob}),
\begin{eqnarray}
P(\tau, {\pmb Q},  \mu) = NC \sum_{r, r'} \sum_{A,B} \int d\mu(k) d\mu(k')
\hat{\rho}_1(k,r;k',r')u_A(k,r) \bar{u}_B(k',r')
\nonumber \\
\times \tilde{\eta}\left(\frac{1}{2}(\omega_k+ \omega_{k'}) \right)\exp\left[ i({\pmb k} - {\pmb k}')\cdot {\pmb Q} - i (\omega_k - \omega_{k'}) \tau \right]
   \left[ \delta_{AB} + \mu \Sigma_{AB}({\pmb Q}, \frac{1}{2} ({\pmb k} + {\pmb k}')) \right]. \label{prob1spin}
\end{eqnarray}

 In deriving Eq. (\ref{prob1spin}), we took the limit $w_{\delta} \rightarrow 1$ and defined as $\Sigma_{AB} $ the Weyl-Wigner transform of the operator $\Sigma$ of Eq. (\ref{tsigma}),
 \begin{eqnarray}
 \Sigma_{AB}({\pmb Q}, {\pmb p}) = \int d^3y e^{i{\pmb y} \cdot{\pmb p}} ( {\pmb Q} - \frac{\pmb y}{2}, B|\hat{T}(\mu)|{\pmb Q} + \frac{\pmb y}{2}, A)
 \end{eqnarray}

We consider the set-up of the ideal time-of-arrival  measurement of Sec. 6.2. That is, we consider an initial state at the origin of coordinates and a detector localized  at ${\pmb x}  = {\pmb Q}$, so that we can treat the system as one-dimensional. Treating the position coordinate $Q$ as a parameter, and we normalize so that $ \sum_{\mu} P(\tau, {\pmb Q},  \mu)$ coincides with the probability density (\ref{toa2}). We obtain

\begin{eqnarray}
P(\tau, Q,  \mu) = \frac{1}{4m} \sum_{r, r'} \sum_{A,B} \int d\mu(k) d\mu(k')
\hat{\rho}_1(k,r;k',r')u_A(k,r) \bar{u}_B(k',r')
\nonumber \\
\times \sqrt{(\omega_k + \omega_{k'})^2 -4m^2} \exp\left[ i(k - k') Q - i (\omega_k - \omega_{k'}) \tau \right]
   \left[ \delta_{AB} + \mu \Sigma_{AB}( Q, \frac{1}{2} ( k +  k')) \right]. \label{prob1spin}
\end{eqnarray}

We integrate over $\tau$ to obtain a probability distribution over $\mu$, $P(\mu)$. Extending the range of integration for $\tau$ to the whole real axis, we obtain
\begin{eqnarray}
P(\mu) = \frac{1}{2}\left[ 1 + \mu  \sum_{r,r'} \int d \mu(k) \hat{\rho}_1(k,r;k,r')\Sigma_{rr'}(Q,k)
 \right], \label{probspin5}
\end{eqnarray}
where
\begin{eqnarray}
\Sigma_{rr'}(Q, k) = \frac{1}{2m} \sum_{AB} u_A(k,r) \bar{u}_B(k,r') \Sigma_{AB}(Q, k).
\end{eqnarray}

We note that if the Wigner-Weyl transform $\Sigma_{AB}$ is independent of $Q$,  the probabilities (\ref{probspin5}) are also independent of $Q$. However, this property is not always guaranteed: the determination of spin depends on a split of the generator of spatial rotations into an angular momentum and a spin part. Since angular momentum depends explicitly on the position coordinate,  it is expected that the spin operator will also depend on $Q$. Our analysis strongly suggests that the spin operators are contextual, i.e., they depend on the specific set-up through which spin is measured. Further analysis is needed in order to demonstrate this point and it will be a subject of a different publication.

 \section{Conclusions}
We have elaborated the formalism for relativistic quantum measurements developed in Ref. \cite{AnSav15a}. We constructed a general  model for a detector using the master equation approximation for the pointer variables, and we identified the mathematical objects necessary for the characterization of a detector. We obtained a simplified expression for the probability associated to $n$ measurement events, and we identified regimes that correspond to ideal measurements.

We applied our results to three classes of systems. In particle detectors, we constructed an ideal probability distribution for time-of-arrival measurements and showed that it persists for different types of field-apparatus coupling. Then, we constructed the probabilities associated to  photodetection events. We showed that Glauber's photodetection theory corresponds to the limit of an incoherent detector. Finally, we considered the case of relativistic spin measurements, where we identified a general expression for the associated probabilities involving an integral over all possible times of detection.

In the last section of Ref. \cite{AnSav15a}, we mentioned some  possible applications of our formalism in relation to the quantum foundations. In addition to this, the elaboration of our results on spin measurements and the consideration of photon polarization measurements are direct extensions of the work presented here.  The strength of this formalism lies in its treatment of multiple measurements. The examples that were presented in this paper involved only single measurement events, but the extension to $n$ events is straightforward. Of immediate  interest are (i) the consideration of time-of-arrival correlations in multi-partite systems and their relation to entanglement, and (ii) the possibility of using the multi-event probabilities in order to provide a Poincar\'e invariant definition of quantum information concepts, in particular, of qubits  and of entanglement.


\end{document}